# Room temperature energy-efficient spin-orbit torque switching in wafer-scale all-vdW heterostructure

Haiyu Wang[1,3,7], Hao Wu[2,7], Yingjie Liu[1], Dongdong Chen[4], Chandan Pandey[1], Jialiang Yin[1], Dahai Wei[4], Na Lei[1], Jie Zhang[1], Haichang Lu[1], Shuyuan Shi[1], Peng Li[5], Albert Fert[1,6], Kang L. Wang[2], Tianxiao Nie[1*], Weisheng Zhao[1*]

[1]*Fert Beijing Institute, MIIT Key Laboratory of Spintronics, School of Integrated Circuit Science and Engineering, Beihang University, Beijing, China.*
[2]*Department of Electrical and Computer Engineering, and Department of Physics and Astronomy, University of California, Los Angeles, California, USA.*
[3]*Shenyuan Honors College, Beihang University, Beijing, China.*
[4]*State Key Laboratory of Superlattices and Microstructures, Institute of Semiconductors, Chinese Academy of Sciences, P.O. Box 912, Beijing, China.*
[5]*Department of Electrical and Computer Engineering, Auburn University, Alabama, USA.*
[6]*Unité Mixte de Physique, CNRS, Thales, Univ. Paris-Sud, University of Paris-Saclay, Palaiseau, France.*

[7]These authors contributed equally: Haiyu Wang, Hao Wu.
[*]Corresponding authors. E-mail: nietianxiao@buaa.edu.cn; weisheng.zhao@buaa.edu.cn

**Abstract:**

The emergent two-dimensional (2D) ferromagnetic materials with unique magnetic properties have endowed great potential for next-generation spintronic devices with extraordinary merits of high flexibility, easy controllability, and high heretointegrability, which is expected to promote the development of Moore's Law continuously. However, it is extremely challenging to realize magnetic switching with ultra-low power consumption at room temperature. Here, we demonstrate the room-temperature spin-orbit torque (SOT) driven magnetization switching in a well-epitaxial all-van der Waals (vdW) heterostructure. The topological insulator $Bi_2Te_3$ not only helps to elevate the Curie temperature of $Fe_3GeTe_2$ (FGT) through interfacial exchange coupling but also works as a spin current source allowing to switch FGT at a low current density of ~$2.2\times10^6$ A/cm$^2$. A large SOT efficiency of ~0.7 is measured at room temperature, and the thickness of FGT is further adjusted to reduce the influence of the thermal contribution on the second-harmonic signal. Furthermore, the temperature and thickness-dependent SOT efficiency prove that the large SOT in our system mainly originates from the nontrivial origin of topological materials. Our experiment has enabled an all-vdW SOT structure and lays a solid foundation for the implementation of room-temperature all-vdW spintronic devices in the future.

**Introduction:**

Spin-transfer torque[1–5] magnetic random access memory (STT-MRAM) is becoming an appealing alternative to overcome the performance bottleneck encountered in the traditional semiconductor-based memory, which could provide much better performances of nonvolatility, high density, and low-power dissipation. However, the initiative parallel or antiparallel collinear magnetic configuration would lead to an incubation delay when using STT for magnetic switching, and the large writing current could break down the tunneling barrier. In comparison, the emerging spin-orbit torque (SOT)[6–10] could eliminate such performance drawbacks and demonstrate much faster speed, better endurance, and higher energy efficiency. Therefore, it is of fundamental and technical importance to use SOT for switching the magnetization, which is expected to become the major competitor for next-generation memories[11–15].

To date, great efforts have been devoted to exploring new principles and materials for realizing high-performance SOT devices[16–20]. Usually, heavy metals that include W, Ta, Pt, *etc.* were employed as the spin current sources through charge-spin conversion, which could exert a torque to the adjacent ferromagnetic layer for magnetization switching[21–24]. For higher SOT efficiency, van der Waal (vdW) topological insulators (TIs) are proposed recently to replace heavy metals due to their unique feature of spin-momentum locking in the non-trivial topological surface state (TSS), which has demonstrated a high-efficiency SOT-driven magnetic switching in three-dimensional (3D) ferromagnet at room temperature with low critical switching current[17,25–28]. However, 3D ferromagnet would limit the size scaling and lower the spin transparency due to the dangling bonding interface[29]. Therefore, it is urgent to develop new material systems with lower dimensions and more superior interface for higher SOT efficiency, which may bring new opportunities to break the power consumption bottleneck of integrated circuits[30].

The recently-discovered vdW 2D ferromagnetic materials could contribute to an atomic flat surface and maintain the magnetic property at 2D limit thickness, which would satisfy such demand. The early Mermin-Wagner-Hohenberg (MWH) theorem predicted that thermal fluctuations in a 2D magnetic system[31,32] forbade the long-range magnetic order at any finite temperature because the continuous symmetry could not be spontaneously broken in a 2D system. However, recently it has been discovered that 2D intrinsic ferromagnetic materials could exist through breaking the MWH theorem by magnetic anisotropy such as $Fe_3GeTe_2$ (FGT) [33], $CrI_3$[34], and $Cr_2Ge_2Te_6$[35], among others. Among them, FGT has received extensive attention because of hard magnetic properties, Kondo lattice behavior, itinerant ferromagnetism, and other fascinating characteristics[36–40]. More interestingly, relevant work has found that intercalating lithium ions into the interlayer gap of FGT can change the density of states on the Fermi surface and successfully raise the $T_c$ to room temperature[41]. Furthermore, our molecular beam epitaxy (MBE)-grown $Bi_2Te_3$/FGT heterostructures were also found to exhibit a $T_c$ above room temperature through interfacial engineering effect mediated by topological insulator $Bi_2Te_3$[30]. These inspiring results promote the FGT as an ideal 2D candidate for exploring SOT-driven-magnetic switching. Recently, the SOT-driven magnetization switching of FGT has been demonstrated using Pt as a spin current source through spin Hall effect or interfacial Rashba-Edelstein effect[42,43]. However, these devices only work at low temperatures (< 200 K) and the challenge is from lacking room-temperature 2D ferromagnetic materials. Furthermore, much higher SOT efficiency could be envisaged through constructing the all-vdW heterostructure, which can provide a clean interface and thus support high interfacial spin transparency. Therefore, there is an urgent need to design all-vdW heterostructures to achieve energy-efficient SOT switching that can operate at room temperature for future 2D spintronic applications.

Here, we realize SOT-driven magnetic switching in the all-vdW MBE-grown $Bi_2Te_3$/FGT heterostructure at room temperature. The SOT-induced magnetization switching is achieved with a critical switching current density of $2.2\times10^6$ A/cm$^2$. The damping-like SOT efficiency was calculated to be about ~0.7 at room temperature. The high efficiency proves the superior characteristics of all-vdW heterostructures constructed from 2D ferromagnetic materials. In particular, the weak vdW interactions between adjacent layers make it possible to combine atomic layers with different matching degrees, thereby getting rid of lattice matching and compatibility restrictions. The high-quality heterostructure interface is one of the most important factors to constitute the high spin transmissivity. Our results provide a paradigm for the construction of all-vdW SOT devices at room temperature and promote the development of 2D ferromagnets for practical applications.

**Magneto-transport measurements in $Bi_2Te_3$, FGT, and $Bi_2Te_3$/FGT heterostructures**

In this work, we deposited thin films on a (0001) sapphire substrate by MBE, combined with the reflection high-energy electron diffraction (RHEED) to in-situ monitor the surface structure of the film during the preparation, and analyzed the surface morphology by atomic force microscopy (AFM). When preparing the wafer-scale all-vdW heterostructure, it is very critical to maintaining the surface flatness of the bottom layer to ensure the optimal lattice matching and compatibility of the two layers. Therefore, after growing topological insulators on (0001) sapphire substrates, the growth temperature needs to be slowly increased in the growth chamber to maintain a Te-rich environment, which will ensure an excellent single crystallinity of the heterostructure. To better understand the sample quality, RHEED was in-situ rotated during the growth process to check the in-plane crystallinity. During the rotation, RHEED stripes changed regularly and coherently, which could exclude the presence of multidomain. To prevent its degradation, we covered the top surface of FGT with a protective layer. Micrometer-sized Hall-bar devices were fabricated by the standard photolithography combined with ion beam etching. The schematic of the device and the measurement setup are shown in Fig. 1a. The Hall-bar structure was patterned with the dimensions of 100 μm (length)×30 μm (width) for electrical transport measurements as shown in Fig. 1b, where $V_{xy}$ and $V_{xx}$ represent the Hall and longitudinal voltage, respectively. As an emergent quantum matter, TIs attract a lot of interest due to the bulk gap and the spin-momentum-locked Dirac fermions on the surface. Hence, for this type of materials such as $Bi_2Te_3$ and $Bi_2Se_3$, it has been proved by both the theory and experiment that surface states consist of a single Dirac cone at the Γ point, and its simplicity has become an ideal object for studying the spintronics and electronics simultaneously. In the following, we grew 8 nm $Bi_2Te_3$ on the sapphire substrate and performed the magneto-transport measurement. By applying an out-of-plane magnetic field, the Hall resistance shows a negative slope, as shown in Fig. 1c, which features the n-type $Bi_2Te_3$. The right inset in Fig. 1c presents the temperature-dependent 2D carrier density ($n_{2D}$)[44] obtained from the Hall data, which reflects the conduction dominantly from the bulk state. The left inset of Fig. 1c elucidates the band structure of $Bi_2Te_3$, and the position of the Fermi level ($E_F$) determines the spin-momentum locking properties. As a layered vdW crystal, FGT has metallic ferromagnetism[33,38]. Each vdW layer is composed of five atomic sublayers with the lattice constants of a = b = 3.9536 (7) Å and c = 16.396 (2) Å. During the growth of FGT in the MBE chamber, the stripe-like RHEED pattern was captured, reflecting an atomic smooth interface, and its crystal structure was further characterized by XRD (Supplementary Fig. S1). To clarify the magnetic behavior of FGT, we conducted magneto-transport measurements by a physical property measurement system (PPMS) on 30 nm FGT thin film. Figure 1d clearly shows the temperature-driven transition from ferromagnetic to paramagnetic state with the $T_c$ around 220 K, which is the same as that in the previous report[41]. Our previous work has shown that the $T_c$ of FGT can be above room temperature by coupling $Bi_2Te_3$ and FGT together based on the interfacial engineering effect[30]. The fascinating phenomenon of the combination of topological insulators and magnetic 2D materials lays the foundation of our current research. Figure 1e shows the hysteresis loops between 80 K and 200 K in $Bi_2Te_3$/FGT heterostructure with an in-plane magnetic field, which verified its PMA feature. Furthermore, the saturation magnetization ($M_s$) and the magnetic properties in $Bi_2Te_3$/FGT were characterized by a superconducting quantum interference device (SQUID) in Fig. 1f, clearly showing the room temperature ferromagnetism. The high-angle annular dark-field scanning transmission electron microscopy (HAADF-STEM) in the inset further proves the atomic structure of $Bi_2Te_3$ and FGT.

**Current-induced SOT switching in $Bi_2Te_3$/FGT heterostructure**

Here, $Bi_2Te_3$(8)/FGT(3) was taken as the research object for SOT switching, where the number in parentheses denotes layer thickness in the unit of nm. Figure 2a shows the geometric diagram of SOT-driven magnetic switching dynamics in the vdW heterostructure of $Bi_2Te_3$ and FGT. By injecting sufficient spin current density ($J_{spin}$) from $Bi_2Te_3$, SOT enables the magnetization (M) switching in the adjacent ferromagnetic layer above the critical writing current density ($J_{write}$). It is worth noting that the injected $J_{write}$ is orthogonal to the accumulated spin polarization direction and the generated $J_{spin}$ direction. Here, $J_{write}$ can be determined as $J_{write} = I_{write}/\left[w * \left(t_{Bi_2Te_3} + t_{FGT}\right)\right]$, where w = 30μm is the width of the Hall-bar[25]. The effective spin-orbit field ($H_{so}$) induced by the spin current is along the tangential of magnetization and could tilt M up or down to get the positive or negative z component ($M_z$). Usually, for PMA samples, an additional external magnetic field ($H_{est}$) needs to be applied during the measurement process to break the mirror symmetry for the deterministic SOT switching. Thus, when sweeping the applied charge current with an external in-plane field, the SOT from the charge-spin conversion in $Bi_2Te_3$ would induce the magnetization reversal in the ferromagnetic layer.

To demonstrate the SOT switching in the $Bi_2Te_3$/FGT heterostructure, a series of in-plane magnetic fields were applied with a 10-ms pulse current along the Hall bar to obtain a deterministic switch polarity. The role of applying an in-plane magnetic field is to overcome the domain wall chirality caused by the Dzyaloshinskii-Moriya interaction (DMI) and thus to realize the deterministic magnetization switching[45]. We observe that the magnetization changes steadily as the applied $J_{\mathrm{write}}$ increases and a complete magnetic switching is achieved when the $J_{\mathrm{write}}$ reaches approximately $4\times10^6$ A/cm$^2$ at 200 K, as shown in Fig. 2b. When the current density is greater than $2.5\times10^6$ A/cm$^2$, Hall-resistance ($R_{\mathrm{H}}$) begins to decrease after reaching maximum, which is related to a Joule heating effect. Our explanation for this case is that the magnetic interactions of FGT are not sufficient to fight against the thermal fluctuations, resulting in a decrease of M[43]. Interestingly, the critical $J_{\mathrm{write}}$ for SOT switching is much smaller than the values reported in FGT/Pt heterostructure probably due to the high efficiency of charge-spin conversion in TI non-trivial origin. As the applied magnetic field is reversed, the opposite chirality of the SOT switching curve demonstrates the typical characteristics of SOT in the PMA sample, as shown in Fig. 2c. Moreover, the device was measured at different temperatures of 210 K and 190 K. We found that as the temperature decreases, the range of the applied magnetic field to achieve the SOT switch gradually increases (Supplementary Fig. S2 and Fig. S3). For an in-depth understanding of the switching behavior, we summarize the dependence of the current density on the applied magnetic field for SOT switching at different temperatures in the phase diagram of Fig. 2d. Here, the critical switching current density ($J_{\mathrm{sw}}$) that is defined as the sign change in $R_{\mathrm{H}}$ is gradually reduced at the higher magnetic field. The deterministic switching happens in the large field and current region, while both up and down magnetization states are possible in the middle of the panel with a small field and current region. Additionally, the switching current decreases with increasing temperature, which is attributed to the simultaneous decrease in $M_s$ as already proved in Fig. 1f.

**Harmonic Hall measurements in $Bi_2Te_3$/FGT heterostructure**

To quantitatively evaluate the SOT efficiency, we use the harmonic Hall measurement to characterize the effective field of SOT, which could provide a solid understanding of each SOT torque, as well as its influencing factors. We apply a small sinusoidal current ($J_{a.c.}$) to the channel of the device and then generates a SOT in the ferromagnetic layer, which will be decomposed into two mutually orthogonal vector components: damping-like torque $\tau_{DL} \sim m \times (\sigma \times m)$ and field-like torque $\tau_{FL} \sim \sigma \times m$ [46]. In the measurement, the frequency is fixed at 133.33 $H_z$ through the lock-in amplification, and the magnetization oscillation of M around the equilibrium position generates the harmonic Hall signals including the in-phase first harmonic Hall voltage ($V_{1\omega}$) and out-of-phase second harmonic Hall voltage ($V_{2\omega}$). We analyzed the second-harmonic anomalous Hall resistance ($R_{AHE}^{2\omega}$) and planar Hall resistance ($R_{PHE}^{2\omega}$) to determine the current-induced SOT effective field. After applying an external magnetic field ($H_x$) to the $x$-plane, the second-harmonic Hall resistance $R_{xy}^{2\omega}$ could be obtained by the following equation[27]:

$$R_{xy}^{2\omega} = R_{AHE}^{2\omega} + R_{PHE}^{2\omega} + R_{ANE}\frac{H_x}{|H_x|} + R_{offset}$$
$$= \frac{R_A}{2}\frac{H_{DL}}{|H_x| - H_k} + R_p\frac{H_{FL}}{|H_x|} + R_{ANE}\frac{H_x}{|H_x|} + R_{offset} \quad (1)$$

where $H_k$ is the magnetic anisotropy field, which is smaller than $H_x$. $H_{DL}$ and $H_{FL}$ are the damping-like effective field ($\propto m \times \sigma$) and field-like effective field ($\propto \sigma$), respectively. $R_p$ and $R_A$ are the planar Hall resistance and anomalous Hall resistance, respectively. $R_{offset}$ is the resistance offset. $R_{ANE}$ is the transverse resistance contributed by the anomalous Nernst effect and other spin-related thermoelectric effects[47]. For the damping-like effective term, it decreases as the external field increases. For the thermal-related term, its sign changes as the external field direction reverse, while its magnitude keeps constant. Usually, the $R_p$ is extremely small compared to the anomalous Hall counterpart, and thus $R_{xy}^{2\omega}$ mainly originates from the damping-like effective field term and thermal-effect term. Figure 3a displays a series of $R_{xy}^{2\omega} - H_x$ curves under different applied $J_{a.c}$ at 200 K. It demonstrates a distinct field dependence, while a step function could also be observed, which means that in addition to the contribution of the damping-like Hall signal, it also has a thermal contribution in our sample. With increasing the $J_{a.c.}$, both signals are enhanced. The inset in Fig. 3a schematically illustrates the second harmonic Hall signal that comes from the SOT-induced magnetization oscillation around the equilibrium position. For quantitatively characterizing the thermal signal, we carried out the temperature-dependent $R_{xy}^{2\omega} - H_x$ at a fixed $J_{a.c.}$ to provide further evidence. Here, we defined $R_{ANE} = (R_{xy(sat_max)} - R_{xy(sat_min)})/2$ to express the thermal contribution, where $R_{xy(sat_max)}$ and $R_{xy(sat_min)}$ are defined as the maximum and minimum values of second-harmonic Hall resistance under a saturated magnetic field[27]. As temperature decreases, the $R_{ANE}$ becomes much larger, which implies thermal contribution is more pronounced at low temperatures. To understand the origin of the thermal-related effect, it is worth noting that the metallic and topological nature of FGT could cause a large anomalous Nernst effect (ANE)[40]. In our sample, the top layer above FGT is air with ambient temperature, while the bottom layer is $Bi_2Te_3$ with a large current. The vertical thermal gradient from the asymmetric structure may contribute to the thermal current, thus inducing the ANE. Nevertheless, we could differentiate the ANE and the SOT-induced second-harmonic Hall resistance through their magnetic field dependence. The right inset of Fig. 3b displays the influence of the FGT's large ANE on the heterostructure and the left inset is a schematic diagram of the step function of the thermal contribution[48].

Relying on the above analysis, we extract $H_{DL}$ and display the dependence of $H_{DL}$ on corresponding $J_{a.c.}$ at 200 K, as shown in Fig. 3c. The resistivities of the $Bi_2Te_3$ and FGT layers are evaluated to be 2490 μΩ cm and 48.5 μΩ cm, respectively (Supplementary Fig. S4). By fitting the process in the large in-plane magnetization region with the formula (1), $H_{DL}/J_{write}$ is ~160.2 Oe per MA/cm$^2$ in the inset of Fig. 3c. The SOT efficiency ($\xi_{DL}$) can be obtained using[49],

$$\xi_{DL} = \frac{2eM_S t}{\hbar}\frac{H_{DL}}{J_{a.c.}} \quad (2)$$

where $e$ and $\hbar$ are the electron charge and reduced Plank constant respectively, $t$ represents the ferromagnetic layer thickness. Accordingly, the value of the $\xi_{DL}$ is determined to be ~5.3 in $Bi_2Te_3$(8)/FGT(3) structure at 200 K.

To eliminate the influence of thermal contribution caused by the ANE of FGT on the SOT efficiency, we adjusted the thickness of FGT to manipulate the shunting current in the $Bi_2Te_3$ for lowering the thermal gradient in the $Bi_2Te_3$/FGT heterostructure[50]. Moreover, we conduct the measurements with different $I_{dc}$ while sweeping $H_x$ to observe the variation of $R_{xy}$, and find that the DC of 0.5 mA to 1.5 mA has no significant effect on the heterostructure, which further verifies that this thickness of the heterostructure has better thermal stability (Supplementary Fig. S5). Figures 4a and 4b display the out-of-plane external magnetic field-dependent $R_{xy}$ on $Bi_2Te_3$/FGT heterostructures with different thicknesses of FGT, including 3 nm and 4 nm at 100 K, 150 K, and 200 K. We normalize its $R_{xy}$ to facilitate comparison. It is worth noting that the $R_{xy}$ of $Bi_2Te_3$(8)/FGT(3) has an obviously negative ordinary Hall slope in the saturated magnetic field region, which is almost the same as that from the $Bi_2Te_3$ Hall signal, indicating that $Bi_2Te_3$ in the heterostructure has a large shunting effect. In comparison, the $R_{xy}$ of $Bi_2Te_3$(8)/FGT(4) shows only the anomalous Hall signal from FGT, which well proves the shunting effect in $Bi_2Te_3$ has been significantly reduced due to more conducting in FGT after increased thickness. The PMA feature was further verified by performing first-harmonic Hall measurement with an in-plane magnetic field and the results under different temperatures are shown in Fig. 4c. Subsequently, we conducted the second-harmonic Hall measurements and display $R_{2\omega}$ signals as a function of $H_x$ under different $J_{write}$ in Fig. 4d. Interestingly, the step function from ANE disappears, which well matches our above prediction. Followed by equations (1) and (2), the room temperature $\xi_{DL}$ is estimated to be ~0.7, which reflects the strong SOC characteristics of TI at room temperature.

To verify the conjecture and understand the related mechanism in our sample, we give a systematic discussion about the temperature dependence of $\xi_{DL}$. Unlike traditional heavy metals, TI exhibits a topologically-protected non-trivial surface state, which is composed of a single massless Dirac fermion with two spin-splitting bands on the surface. When the time-reversal symmetry is broken, the surface state will open a gap. The bulk Hamiltonian projected onto the surface state is described as[51,52]:

$$H_{surf}\left(\overrightarrow{k_x}, \overrightarrow{k_y}\right) = \upsilon\hbar\left(\overrightarrow{\sigma^x}\,\overrightarrow{k_y} - \overrightarrow{\sigma^y}\,\overrightarrow{k_x}\right) \tag{3}$$

where $\upsilon$ is the velocity of the surface state and $k$ is the Dirac electron momentum. When the $J_{write}$ is applied to TI, the spin of the Dirac electron is locked, and the movement of the Fermi surface in the $k$-space will produce controllable spin polarization. Another important origin of SOT is the spin Hall effect (SHE) of the bulk state, the phenomenon of which utilizes the bulk SOC in TI to convert non-polarized write current into the spin current. Due to the asymmetric scattering of conductive electrons, the spin-up and spin-down are deflected in opposite directions, forming a transverse spin current.

Figure 5a displays the schematic spin-related band structure of the TSS and bulk state. Both of them coexist in the film[53], and either the surface or bulk state would provide a contribution to the final SOT. To gain insights on how large surface contribution for SOT, temperature-dependent SOT efficiency and its relation to the $E_F$ were carried out for analysis. Figure 5b shows the precise SOT efficiency results through harmonic Hall measurements in the $Bi_2Te_3$(8)/FGT(4) heterostructure (more details in Supplementary Fig. S6). We found that $\xi_{DL}$ exhibited a drastic nonlinear growth with a decrease in temperature[54]. The $E_F$ occupies a very conductive bulk state at room temperature, and it shifts downwards to Dirac cone with a reduced bulk state as the temperature decreases (Supplementary Fig. S7). The fact that TI with reduced bulk conductance leads to a higher SOT efficiency suggests that the TSS renders significant contributions to the efficient SOT. Furthermore, additional heterostructures with different TI thicknesses ($Bi_2Te_3$(6)/FGT(4) and $Bi_2Te_3$(10)/FGT(4)) were prepared for comparison with previous samples. The SOT efficiency from the harmonic measurements has undergone a dramatic increase, which further proves the significant surface contribution for $Bi_2Te_3$ at room temperature (Supplementary Fig. S8). Besides, the Rashba spin-splitting surface state in the two-dimensional electron gas (2DEG) may coexist with the TSS in the $Bi_2Te_3$ due to the band bending and structural inversion asymmetry[5,54]. However, the Rashba effective magnetic field is expected to increase gradually as the temperature rises in the semiconductor system[55,56], different from our experimental results. Hence, we believe that the Rashba-split surface state is not the main physical mechanism for SOT switching[26].

For the chirality of SOT, the relationship between the $J_{spin}$ and the $J_{write}$ can be expressed by the following formula[57]:

$$J_{spin} = \frac{\hbar}{2e}\theta_{SH}(J_{write} \times \vec{\sigma}) \tag{4}$$

where $\theta_{SH}$ is the spin Hall angle, σ is the polarization of the spin and its direction is orthogonal to the direction of the $J_{write}$. For non-ferromagnetic materials that provide spin currents, the spin direction of the top surface and the bottom surface is opposite and its chirality is defined by the sign of $\theta_{SH}$. Compared with our results, the SOT switching in FGT/Pt heterostructures shows the same chirality, further accurately confirming our conclusion[42,43]. As reported previously, the chirality from TSS is the same as that from the bulk state with positive spin Hall angle[27,58]. Finally, the SOT switching of the FGT layer was successfully demonstrated in the $Bi_2Te_3$(8)|FGT(4) heterostructure at room temperature when 10-ms pulse currents were applied to the Hall bar with an $H_{est} = \pm 2\ kOe$ under several consecutive sweeps, as shown in Fig. 5c. It sets a new stage for exploring all-vdW SOT devices.

For clarity, we summarize the switching write current density and its realized maximum temperature of several representative heterostructures for comprehensively understanding the SOT feature in the $Bi_2Te_3$/FGT heterostructure, and the results are presented in Table 1[43,59–61]. The heavy metal Pt is generally used as the preferred material to achieve SOT switching of FGT at low temperatures. It is worth noting that the minimum $\xi_{DL}$ in the FGT/Pt heterostructure reported by Alghamdi *et al.* is as large as the maximum of the CoFeB/Pt structure[42], demonstrating the vdW FGT superiority. When compared with our sample, the large $\xi_{DL}$ value well proves the TI of $Bi_2Te_3$ is superior for charge-spin conversion even with 2D vdW ferromagnet. In some cases, such as previously-reported $Bi_2Te_3$/Ti/CoFeB/MgO heterostructure[27], the SOT efficiency is nearly an order of magnitude smaller when compared to our sample[27]. Recently $WTe_2$/FGT heterostructures have also been found to achieve SOT properties and relatively excellent performance, but still at low temperature[59,60]. These results obtained with the same characterization method may provide evidence that the interfacial spin transparency could be significantly enhanced by the vdW-gapped interface between $Bi_2Te_3$ and FGT.

In such a case, the $\xi_{DL}$ is related to the internal $\theta_{SH}$ of TI and the interfacial spin transparency $T_{int}$[29,43]. The interfacial spin transparency could be mainly determined by the mechanisms of spin backflow and spin memory loss, which could be characterized by the effective spin-mixing conductance and the spin conductance of the non-ferromagnetic layer[29]. A good interface contributes to the transparency during spin transport at room temperature, which is one of the most important factors to achieve energy-efficient SOT switching in an all-vdW heterostructure and highlights the strong SOC characteristics of TI. In addition to the contribution of the $Bi_2Te_3$, we may consider that there is another SOT contribution to the enhancement of $\xi_{DL}$ due to the $H_{so}$ generated by FGT itself, which further assists the M switching. This can be attributed to the topological nodal-line semimetal feature of FGT, whose topological band structure enhances the intrinsic contribution of Berry curvature to the anomalous Hall effect and $H_{so}$[62,63].

To summarize, wafer-scale vdW $Bi_2Te_3$/FGT heterostructure prepared by MBE has successfully realized room-temperature ferromagnetism and current-driven SOT switching. A series of electrical transport measurements display that the SOT induced by the TI can realize the magnetization switching at room temperature. The harmonic Hall measurements reveal that the SOT efficiency was as high as ~0.7 at room temperature, and this value could be further increased to ~2.69 with decreasing TI thickness. Together with the temperature-dependent measurement, the high charge-to-spin conversion efficiency may be mainly attributed to the improved interfacial spin transparency and nontrivial origin of topological materials from all-vdW $Bi_2Te_3$/FGT structure. The realization of room-temperature ferromagnetism and SOT switching together in $Bi_2Te_3$/FGT heterostructure points to a superior approach for the development of all-vdW heterostructures, and lays the foundation for implementation of room-temperature 2D vdW spintronic devices in the future.

## Methods

**Sample Growth.** The (0001) sapphire substrate was used to grow the sample. High-purity Bi, Fe, Ge, and Te were evaporated from Knudsen effusion cells in the MBE system with the base vacuum of $10^{-10}$ Torr. After degassing at high temperature, the substrate was cooled down to 300 ºC for growing both the FGT thin film and $Bi_2Te_3$/FGT heterostructure with a growth rate of ~0.05 Å/s, and the sample quality was monitored by an in-situ RHEED system.

**Characterization.** The morphologies of the samples were investigated by AFM. The microstructure and composition were comprehensively characterized by XRD and HAADF in STEM mode. The cross-section TEM sample was prepared by a focused ion beam (FIB). MOKE and SQUID were employed to measure their magnetic properties. Furthermore, the magnetotransport studies were carried out in the Quantum Design physical property measurement system (PPMS).

## Acknowledgments

This work was supported by the National Key R&D Program of China (2018YFB0407602), the National Natural Science Foundation of China (61774013), the International Collaboration Project (B16001), and the National Key Technology Program of China (Grant No. 2017ZX01032101).

## Author Contributions

T.X.N. and W.S.Z. conceived the ideas. T.X.N. designed the experiments. H.Y.W. and Y.J.L. contributed to the MBE growth. C.P. and D.D.C. fabricated devices. H.Y.W. performed electrical measurements. T.X.N. and H.Y.W. analyzed the data. T.X.N. H.Y.W. and H.W. wrote the paper. All the authors discussed the results and commented on the manuscript.

## Competing financial interests statement

The authors declare no competing financial interests.

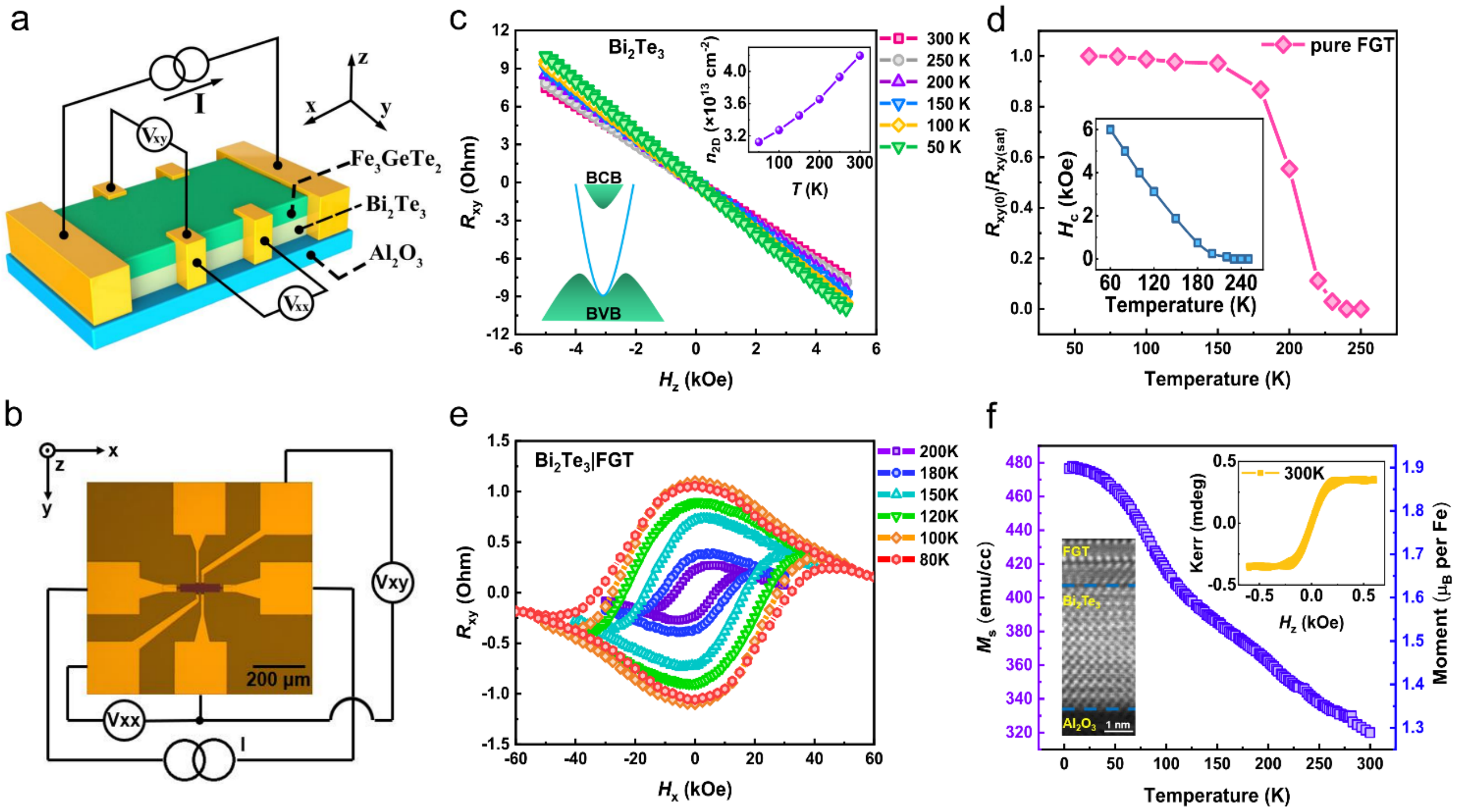

**Figure 1 | Electrical and magnetic measurements in $Bi_2Te_3$, $Fe_3GeTe_2$ and $Bi_2Te_3/Fe_3GeTe_2$. a,** Schematic diagram of the device measurement setup. **b,** Optical micrograph of a Hall bar device for electric measurement. **c,** $R_{xy}$ - $H_z$ curves at different temperatures in 8nm $Bi_2Te_3$. The left inset shows the schematic of the band structure and the right insert shows the temperature-dependent carrier density. **d,** Normalized remnant anomalous Hall resistance, and coercivity as a function of temperature in pure FGT, which displays the Curie temperature is ~220 K. **e,** $R_{xy}$ as a function of in-plane field $H_x$ at

different temperatures in $Bi_2Te_3$(8)/FGT(3) heterostructure, which displays the perpendicular magnetic anisotropy. **f,** Curves of saturation magnetization $M_s$ at different temperatures in $Bi_2Te_3$/FGT heterostructure. The right inset displays the Kerr signal of the heterostructure at 300 K and the left inset displays the crystalline quality by HAADF-STEM image.

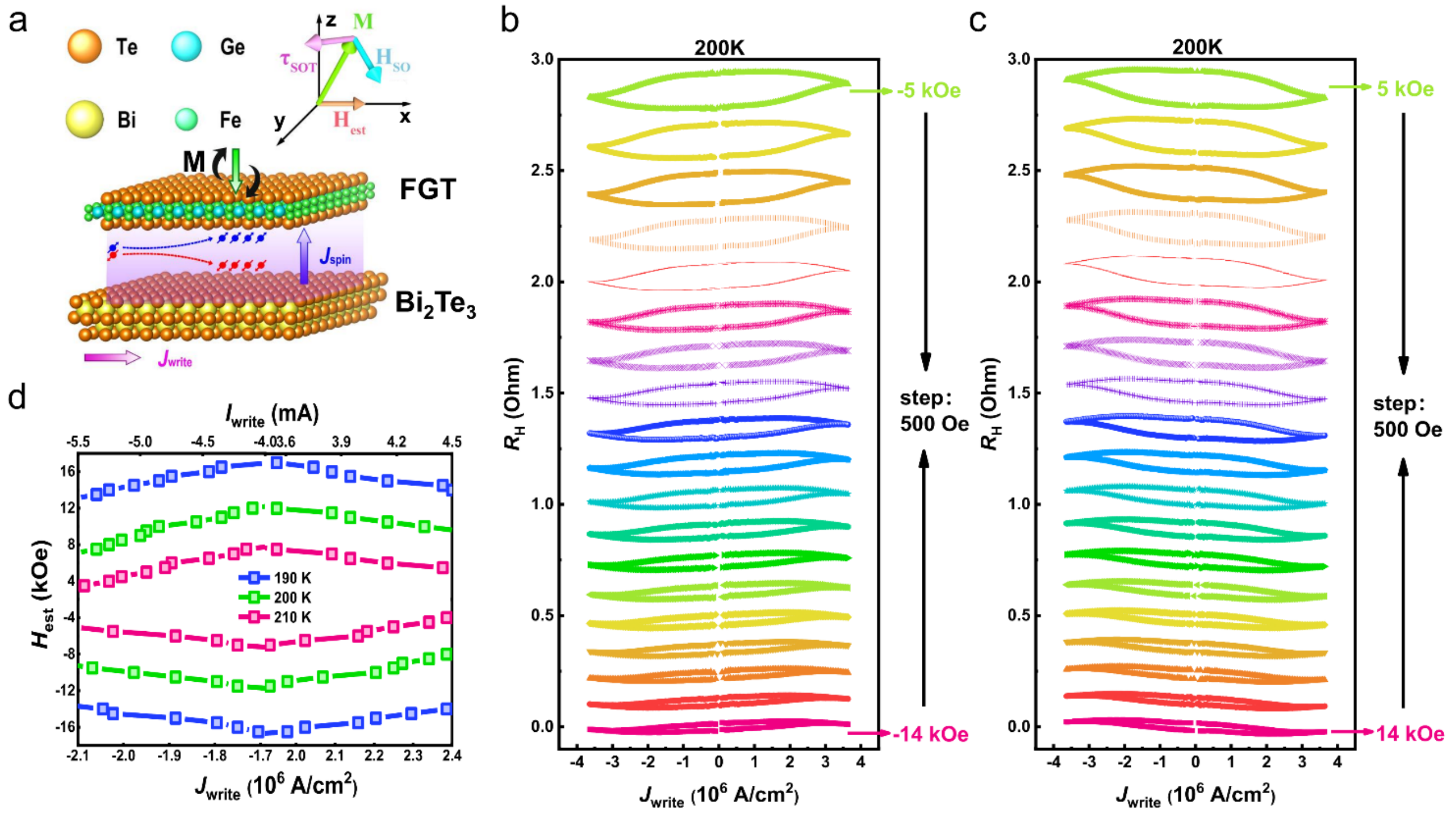

**Figure 2 | Spin-orbit torque induced magnetization switching behaviors in $Bi_2Te_3$(8)/$Fe_3GeTe_2$(3) heterostructure. a,** Geometric structure diagram of SOT switching in $Bi_2Te_3$/FGT heterostructure. The effective spin-orbit field ($H_{so}$) exerts a spin torque ($\tau_{SOT}$) for magnetization switching. **b, c,** Current-induced magnetic switching presented in Hall bar structure at 200 K with different in-plane magnetic field, showing the opposite SOT switching chirality when reversing magnetic field. **d,** Phase diagram for SOT switching in different writhing currents and in-plane magnetic fields at 190 K, 200

K, and 210 K.

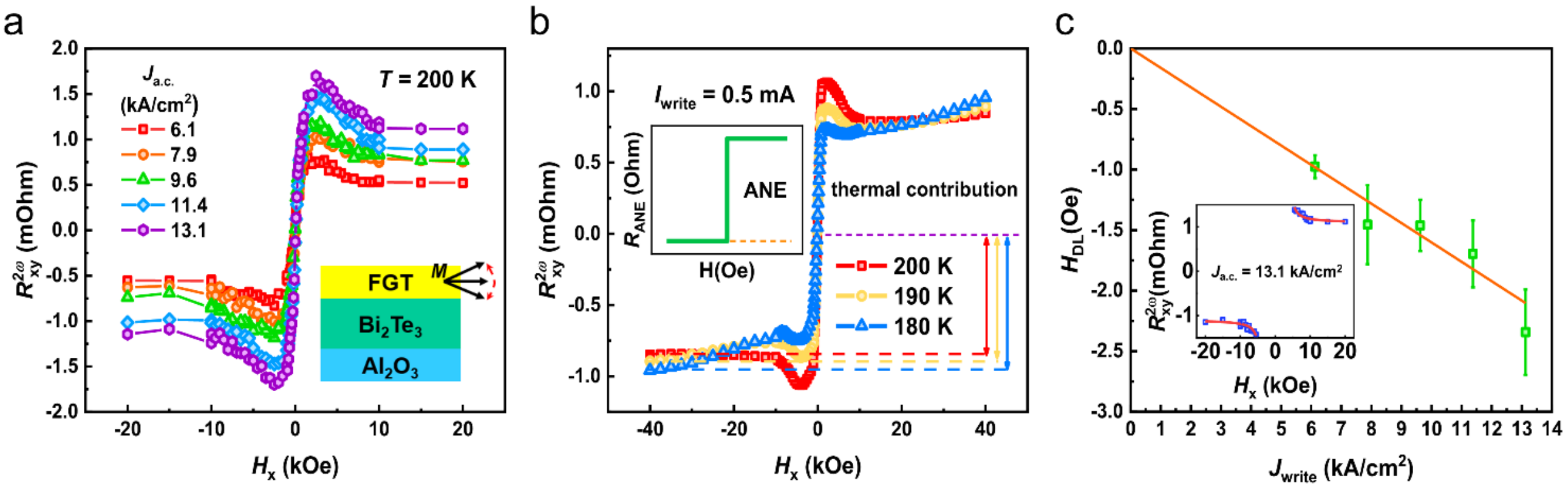

**Figure 3 | Harmonic measurements under different temperatures in $Bi_2Te_3$(8)/$Fe_3GeTe_2$(3) heterostructure. a,** Second-harmonic Hall resistance ($R_{xy}^{2\omega}$) under different $J_{a.c.}$ at 200K. The inset displays the oscillation of magnetic moment at equilibrium position in harmonic measurement. **b,** Second-harmonic Hall resistance ($R_{xy}^{2\omega}$) as a function of in-plane magnetic field ($H_x$) at different temperatures under a constant write current density. The inset displays the field dependencies of anomalous Nernst resistance. **c,** Damping-like effective field ($H_{DL}$) as a function of the current density extracted by fitting the second harmonic Hall signal. The inset shows a typical $R_{xy}^{2\omega}$- $H_x$ curve under the large magnetic field range for fitting out the $H_{DL}$.

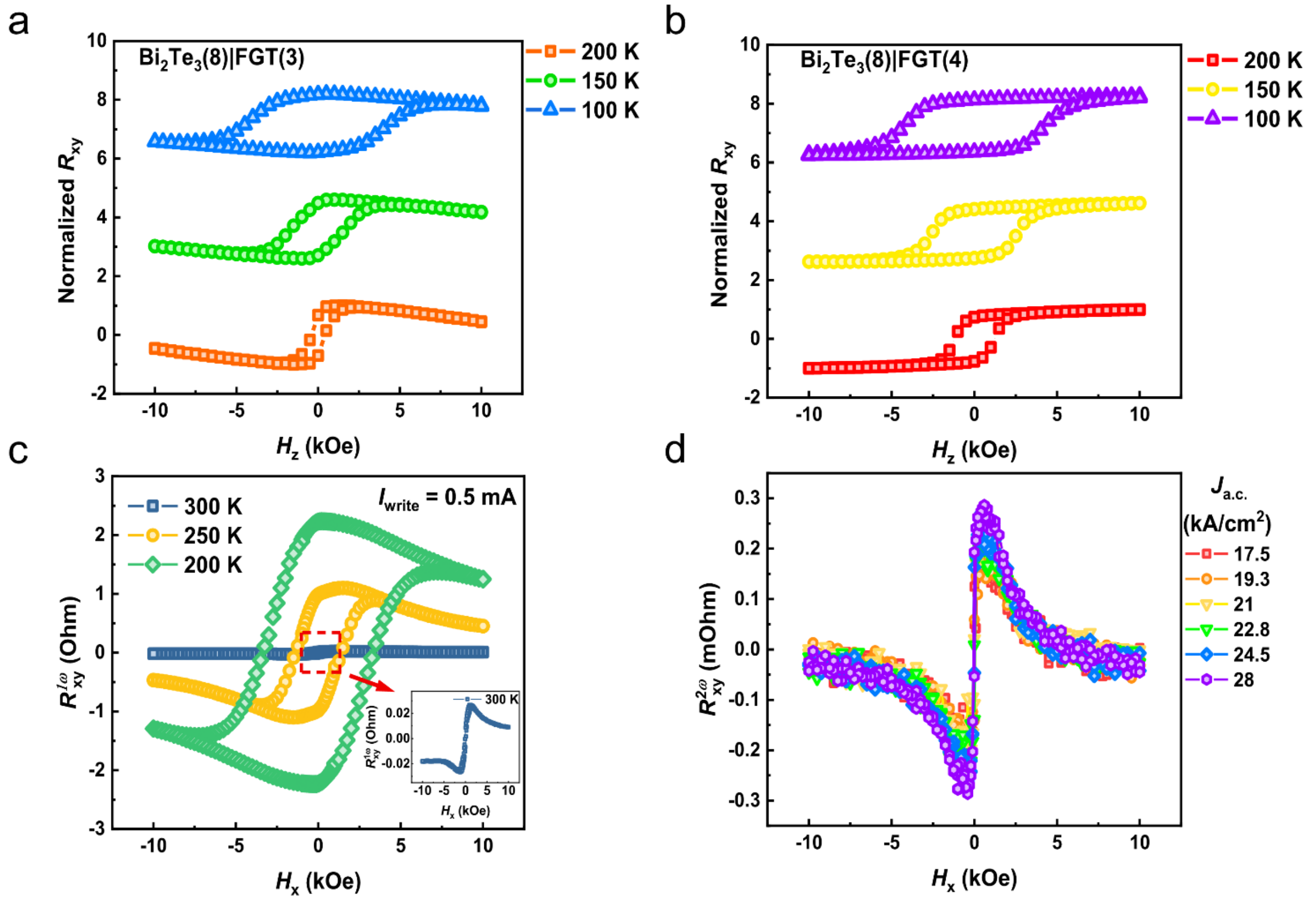

**Figure 4 | Comparative out-of-plane field anomalous Hall results and in-plane field harmonic Hall signals in $Bi_2Te_3$(8)/$Fe_3GeTe_2$(4) heterostructure. a, b,** Normalized anomalous Hall resistance ($R_{xy}$) as a function of the out-of-plane external magnetic field ($H_z$) at 100 K, 150 K and 200 K in $Bi_2Te_3$(8)/$Fe_3GeTe_2$(3) and $Bi_2Te_3$(8)/$Fe_3GeTe_2$(4) heterostructures, respectively. **c,** First-harmonic Hall resistance as a function of in-plane external field under different temperatures in

$Bi_2Te_3$(8)/$Fe_3GeTe_2$(4) heterostructure. **d,** Second-harmonic Hall resistance under different applied $J_{a.c.}$ at room temperature, showing the SOT enhancement with increasing current.

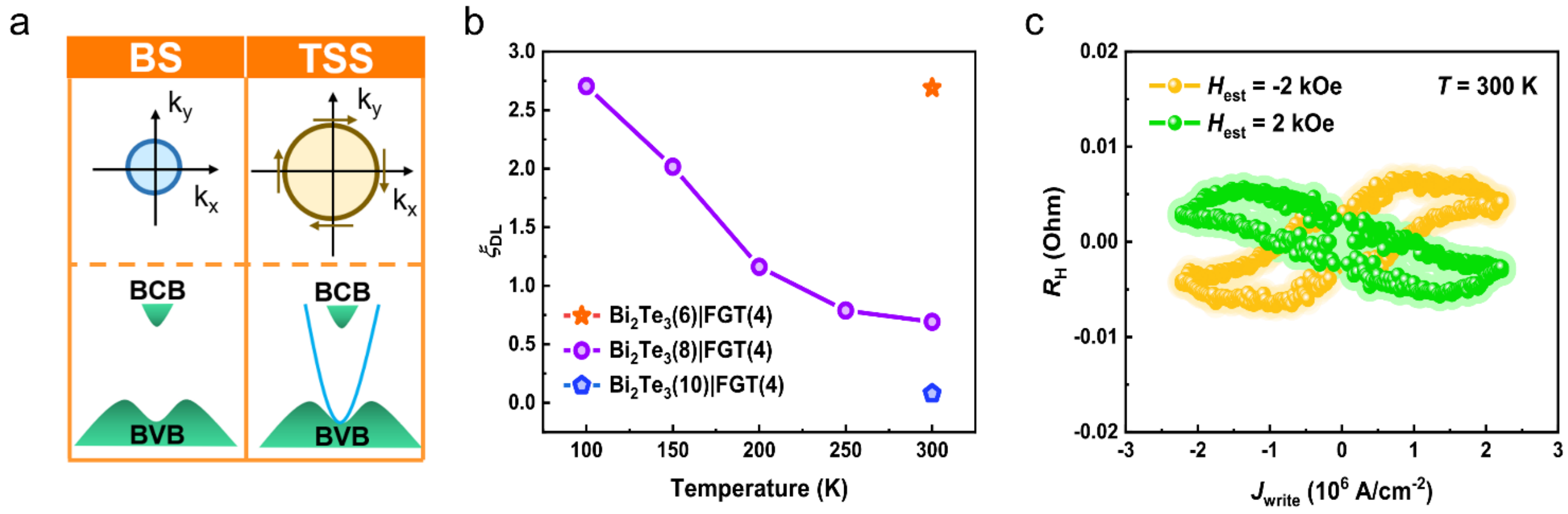

**Figure 5 | SOT efficiency characterization and current-induced room-temperature switching in $Bi_2Te_3$/$Fe_3GeTe_2$(4) heterostructures. a,** Illustration of the concept of charge-to-spin conversion via bulk state and topological surface states. **b,** The SOT efficiency ($\xi_{DL}$) at different temperatures and thickness, showing the enhanced SOT switching. **c,** Current-induced magnetization switching under ±2 kOe at room temperature.

**Table 1 | SOT characteristics in several typical heterostructures of 2D vdW ferromagnets**

| | Maximum Temperature | Switching write current density |
|---|---|---|
| **$Pt|Fe_3GeTe_2$ (ref. [43])** | 120 K | ~$7.4\times10^6$ A/cm$^2$ |
| **$Pt|Fe_3GeTe_2$ (ref. [61])** | 180 K | ~$1.5\times10^7$ A/cm$^2$ |
| **$WTe_2|Fe_3GeTe_2$ (ref. [59])** | 190 K | ~$4.2\times10^6$ A/cm$^2$ |
| **$WTe_2|Fe_3GeTe_2$ (ref. [60])** | 160 K | ~$3.5\times10^6$ A/cm$^2$ |
| **$Bi_2Te_3|Fe_3GeTe_2$ (this work)** | 300 K | ~$2.2\times10^6$ A/cm$^2$ |